\documentclass{azh}

\usepackage{graphicx}
\usepackage{latexsym}

\newcommand{\km}{\,\mbox{km}\,\mbox{s}^{-1}}
\def\Ha{\hbox{H$\alpha$\,}}
\def\Hb{\hbox{H$\beta$\,}}

\begin{document}

\title{The gas emission spectrum in a star-forming region in the BCD galaxy VII~Zw~403 (UGC~6456)}

\author{V.P. Arkhipova$^{1}$ \and T.A. Lozinskaya$^{1}$ \and A.V. Moiseev$^{2}$ \and O.V. Egorov$^{1}$}

\institute{Sternberg Astronomical Institute, Universitetskii pr.
13, Moscow, 119992 Russia \and Special Astrophysical Observatory,
RAS, Nizhnii Arkhyz, Karachai-Cherkessian Republic, 357147 Russia}

\offprints{T.A.  Lozinsksaya, \email{lozinsk@sai.msu.ru}}

\date{}

\titlerunning{The gas emission spectrum in the BCDG VII~Zw~403}

\authorrunning{Arkhipova  et al. }

\abstract{Observations with the 6-m telescope of the Special Astrophysical Observatory obtained with
the MPFS integral-field spectrograph and a longslit spectrograph with the SCORPIO focal reducer are used to
analyze the emission spectrum of the ionized gas in a star-forming region in the BCD galaxy VII~Zw~403.
We present images of the galactic  central region in the \Ha, \Hb, [SII], and [OIII] emission lines, together with maps of the relative [OIII]/\Hb and [SII]/\Ha intensities. We have determined the parameters of the gas in bright ionized supershells, and estimated the relative abundances of oxygen, nitrogen, and sulfur; a low relative N/O abundance was detected.}

\maketitle

\section{Introduction}

Indications of active star formation in blue compact dwarf (BCD)  and Irr galaxies include
giant complexes containing most of the young stellar population, bright HII
regions, and multiple shells and supershells of various sizes. Such complexes
clearly reveal interaction of OB associations with the ambient gas, providing
unique possibility for observational verification of the classic theory of
interaction of collective winds and supernovae with the interstellar medium,
which is so far in a fairly poor agreement with observations.

This paper continues  studies of the interstellar medium in a complex
associated  with a  recent burst of star formation  in the BCD galaxy VII~Zw~403
(UGC~6456) initiated by Lozinskaya et al.~[\cite{1}].

VII~Zw~403 is one of the closest BCD galaxies (at the distance $d=4.5$~kpc),
is isolated and slowly rotating and has undergone  several star-formation
episodes of various  intensities~[2,3]. The most recent burst of star formation
happened 4--10~Myr ago and encompassed  a central region of the  galaxy, about
one kpc in size, in the direction of a giant neutral hydrogen cloud with a very
high column density~[4].

 Several sites associated with this latest star-formation episode are observed; the youngest
massive stars form compact OB associations Nos. 1-6 (for uniformity, like in [1], we use the notation for
associations and HII regions of Lynds et al.~[2]). More evolved stars, including high-luminosity red
supergiants, are distributed in an extended, elliptical area and show no concentration toward the compact
associations. The ionized gas is concentrated in the same central region of the galaxy, and exhibits
signs of several sites of recent star formation, in the form of bright HII regions associated with compact
associations and faint diffuse emission surrounding them [2, 5, 6]. Observations with the HST  have revealed shell-like structures 80-100 pc in size for a number of bright HII regions. Lynds et al. [2] used slit spectrograms obtained on the 4-m telescope of the Kitt Peak National Observatory
to derive the shells' expansion rate, $50-70\km$, although these results were not confirmed by Lozinskaya
et al. [1] (see below). Silich et al. [7] found traces of a giant ($D \simeq 500$~pc), faint ring in \Ha emission produced by diffuse gas. Reports of the presence of an extended region of diffuse X-ray emission inside this giant ring remain, for the moment, unconfirmed [8-12].

In our paper [1], we presented a detailed study of the large-scale structure and kinematics of the
ionized gas using observations obtained on the 6-m telescope of the Special Astrophysical Observatory
(SAO) Russian Academy of Sciences with the SCORPIO multi-mode focal reducer in three modes: direct
imaging (in the \Ha, [OIII], and [SII] lines), longslit  spectroscopy, and spectroscopy with a scanning
scanning Fabry--Perot interferometer.  In addition to the known bright HII regions and traces of the faint giant ring, we detected many new diffuse and ring structures, detected fine structure in the giant ring, and mapped
the radial-velocity distribution at the maximum and half-maximum levels of the \Ha line for the entire
central region of the galaxy. To search for signs of expansion, we plotted so-called `velocity ellipses'
and distributions of the half-width of the \Ha line for all the HII regions; however, we found no variations of
the radial velocity at the line maximum and/or halfwidths providing evidence for expansion of the shells
with a velocity of $50-70\km$, as reported in [2]. Our longslit spectrograph observations in the red (\Ha,
[NII], [SII]) and green  (\Hb, [OIII]) ranges likewise did not reveal any indications of the shells' expansion with such velocities. On the contrary, our Fabry-Perot interferometer and longslit spectrograph data displayed
an obvious line that broadened (from $\textrm{FWHM} \simeq 60{-}70\km$ to $\textrm{FWHM} \simeq
80{-}120\km$) in the region of the faint, diffuse emission outside the bright HII regions.

\begin{figure}[t!]
\includegraphics[scale=1.0]{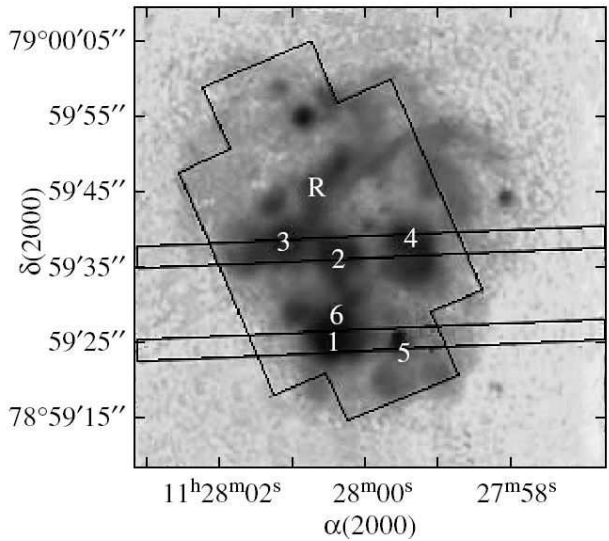}
\caption{Direct H$\alpha$ image of the region of current star
formation from~[1]. The part of the galaxy covered by  the MPFS observations and
the positions of the spectrograph slit are shown. The numbers mark the
bright HII regions noted  in the Introduction (according to the
designations of Lynds et al.~[2]), the letter $R$ marks the brightest part
of the giant ring.}
\end{figure}

The kinematic age of the bright shells corresponding to our estimates of their expansion velocity
($20 \km$ or lower) is at least 2-4 Myr, in good agreement with the ages of the corresponding
compact OB associations estimated in [2]. The extended filamentary and diffuse regions of ionized gas
detected essentially throughout the central part of the galaxy and the giant HII ring could be related to
the older star population from the most recent burst of star formation (with an age of 10 Myr,
according to [2]). We demonstrated in [1] based on our direct images that the large-scale structure of the star-forming region was generally the same in the [OIII], [SII], and
\Ha + [NII] lines. In two extended, very faint, diffuse regions, we found enhanced relative [OIII]/\Ha and
[SII]/\Ha line intensities. However, due to the low brightness of these diffuse regions, these preliminary
conclusions needed to be checked using observations with an integra-field (3D, panoramic) multi-pupil fiber spectrograph. In the brightest shell (No. 1) around the richest and youngest association (No. 1), we detected
in [1] the presence of weak high-velocity wings: in the [OIII] line, out to velocities from $-200$ to $-300 \km$ from the line center for intensities of 5\% of the line maximum, and in the \Ha line, out to $-350\km$ and
$550-600\km$ for intensities of about 2\% of the line maximum. This was the first detection of velocities
this high in this galaxy, and provides clear evidence for acceleration of gas at shock fronts. Panoramic spectrograph observations are of special interest for shell No. 1.

In the present study, we analyze the emission spectrum of the gas in the star-formation region
using observations of the galaxy obtained with the integral-field multi-pupil fiber spectrograph (MPFS)
and longslit spectrograph of the SAO 6-m telescope. We present our observations and reduction techniques in
the next section, followed by the main results of the observations and a discussion and conclusions.

\section{Observations and data reductions}

\subsection{Observations with the MPFS}

Our observations with the integral-field spectrograph MPFS at the  primary focus of the SAO 6-m telescope were made on May 23/24 and 24/25, 2004. A description of the spectrograph is given by Afanasiev et al.~[13]
or at web-address \verb"http://www.sao.ru/hq/lsfvo/devices.html". The detector
was a $2048\times2048$-pixel EEV~42-20 CCD chip. The spectrograph enables the recording of spectra  from 256 spatial elements (designed as square lenses), forming an array of $16\times16$ elements in the sky plane. The scale was $1''$ per lens. Simultaneously, we recorded the night-sky spectrum of a field at a $4'$ distance from the center
of MPFS field of view. The spectra had a resolution of about 8~\AA\ and covered the wavelegth range $4250{-}7200$~\AA. In total, we observed seven fields in the galactic central region, displaced with respect of each other. Table~1 presents the total exposure time, $T_{\textrm{exp}}$, and the mean
seeing for each of the fields.

\begin{table}[t!]

\caption{Log of MPFS spectroscopy.}

\begin{tabular}{c|c|c|c}
\hline
Field & Date & $T_{\textrm{exp}}$, s & Seeing, $''$\\
\hline
1    &2004 May 23/24 &  1800 & 2.0 \\
2    &2004 May 23/24 &  1800 & 2.0 \\
3    &2004 May 24/25 &  1800 & 1.7 \\
4    &2004 May 24/25 &  1800 & 1.7 \\
5    &2004 May 24/25 &  1200 & 1.7 \\
6    &2004 May 24/25 &  1200 & 1.7 \\
7    &2004 May 24/25 &   600 & 1.7 \\
\hline
\end{tabular}
\end{table}

\begin{figure*}[t!]
\includegraphics[scale=1.2]{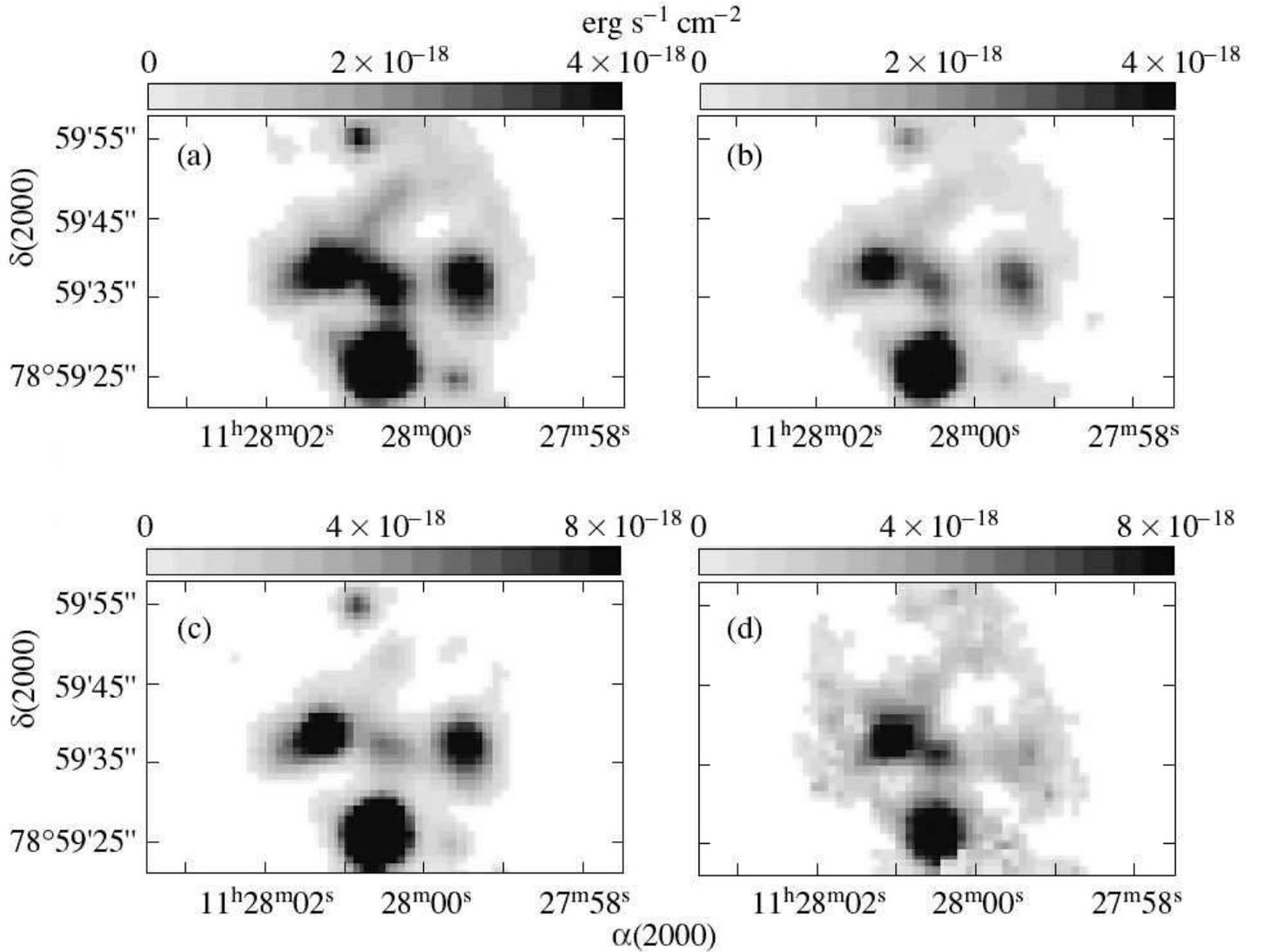}
\caption{Images of the central part of the galaxy in the (a) H$\alpha$,
(b) H$\beta$, (c) [SII]$\lambda6717+6731$\AA~(c), and (d) [OIII]$\lambda5007$\AA\,lines from
 MPFS observations.}
\end{figure*}

We reduced our spectral observations using IDL-based software developed in the SAO's Laboratory of Spectroscopy
and Photometry of Extragalactic Objects by V. Afanasiev and A. Moiseev. The primary reduction
included bias subtraction, flat-fielding, removal of cosmic-ray hits, extraction  of individual spectra
in the CCD images, and wavelength calibration using the spectrum of a He-Ne-Ar calibration lamp.
The night-sky spectrum was subtracted from the linearized spectra. The observed fluxes were then
expressed in an absolute energy scale using spectra of the spectrophotometric standard Grw+70 5824,
which was observed immediately after VII~Zw~403 at a similar zenith distance. We used the mean
atmospheric-extinction spectral curve for the SAO given in [14] to carry out air-mass corrections.

The data reduction yields a data cube, with each $16\times16$ pixel of the image represented by a spectrum
of 2048 elements. After the preliminary reduction, the data cubes for all seven fields were put together
and summed, so that the resulting mosaic had a size of $49''\times31''$; the corresponding region of the
galaxy is shown in Fig. 1.

\begin{figure}[t!]
\includegraphics[scale=1.]{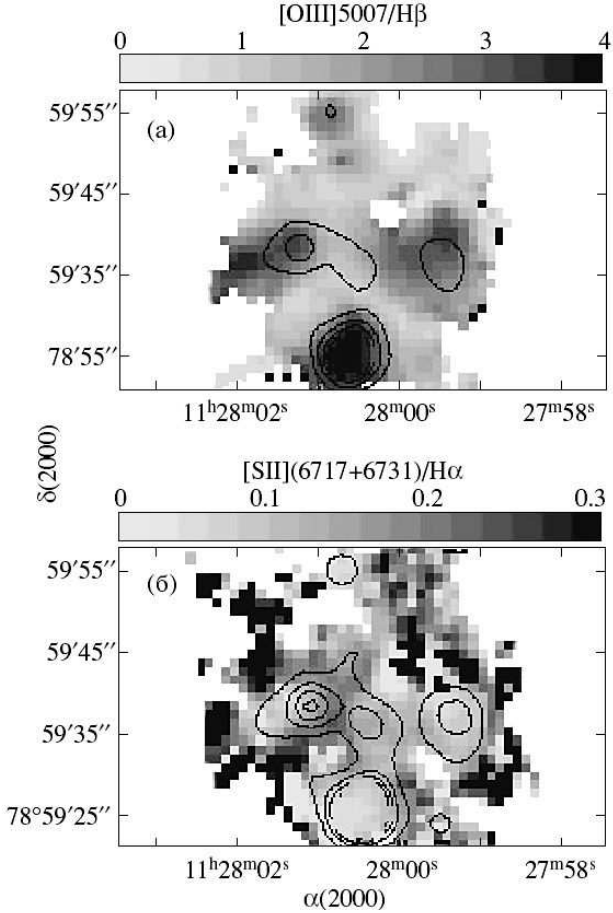}
\caption{Distribution  of the relative line intensities,
for (a) [OIII]$\lambda500$7/H$\beta$ and (b) [SII]$\lambda(6717$+6731)/\Ha. The
contours show the image brightness in the (a) H$\beta$ and (b) H$\alpha$
lines. Darker shades indicate higher intensity ratios.}
\end{figure}

\subsection{Longslit spectra}

In this paper, we use the spectra analyzed in [1] to study the emission characteristics in individual lines.
A log of our observations and a description of the instrument and data-reduction techniques are given
in [1]. Here, we only briefly note that the observations were acquired at the prime focus of the SAO 6-m telescope with the SCORPIO multi-mode instrument [15]. We obtained two spectra (slit positions are shown in
Fig. 1: section 1 passed through HII regions Nos. 2,  3, and 4 and section 2 through regions Nos. 1 and
5. In both cases, the position angle of the spectrograph slit was $88^\circ$. The spectra were recorded in two
spectral ranges, corresponding to $\lambda\lambda4800{-}5600$~\AA\,
which includes the bright H$\beta$ and [OIII] lines, and $\lambda\lambda6270{-}7300$~\AA, which contains the \Ha, [NII], and [SII] lines. The spectral resolution was $2.2{-}2.5$~\AA.

\section{Results}

Figure 1 presents the direct \Ha image of the  star-formation region in the galaxy obtained
in [1]; the numbers correspond to the bright HII regions mentioned in the Introduction. The figure
shows the part of the galaxy covered by our MPFS observations and the positions of the spectrograph
slit. Images of the galactic  central region in the \Ha, \Hb, [SII]$\lambda6717,6731$~\AA, and [OIII]~5007\AA\
lines obtained from  MPFS observations are shown in Fig. 2.

Figure 3 displays maps of the distribution of the relative [OIII]/H$\beta$ and
[OIII]/H$\beta$ line intensities based on these observations, traditionally used for diagnostics
of the emission from HII regions and gas behind shocks. The contours in Figs. 3a and 3b show the
brightness of the images in the \Hb and \Ha lines, respectively.

It follows from Fig. 3 that the preliminary conclusion made in [1] that the relative [OIII]/H$\beta$ and
[OIII]/H$\beta$ line intensities were enhanced inside the giant ring, between bright HII regions Nos. 2 and 4, is not confirmed by our MPFS observations. (This conclusion in paper [1] was based on filter images of the
galaxy in the corresponding lines.)  The second, still fainter, region mentioned in [1] is outside the giant
ring, and is not covered by our MPFS observations.

We determined the mean relative intensities of the `diagnostic' lines in the five bright HII regions in
two ways: from the observations with the MPFS integral-field spectrograph and from the longslit spectrograms.
The line intensities in the bright part of the giant ring were measured only with the integral-field data. The results are given in Table 2, which also contains the data for the bright HII regions obtained
by Lynds et al. [2] with a longslit spectrograph on the 4-m telescope of the Kitt Peak National Observatory.
The longslit  observations in both studies were obtained under similar conditions, and
with the same slit position. The spectral resolution (estimated from sky lines) was $FWHM = 120 \km$
in our study, comparable to that of [2]; the angular resolutions were the same, equal to $2''$ (this was
limited by seeing in our observations, and by the slit width in [2]).

Table 2 contains the formal uncertainties of the means. The actual accuracy of our measurements is
clear from a comparison of the relative intensities in the slit spectra and MPFS observations. Here, the
main cause of disagreement between the estimates obtained using these two methods is the different
integration regions used when deriving the mean line intensities for each object. All the data in Table 2
remain uncorrected for extinction.

Table 2 presents the relative intensities for the
[OIII] line for the central region and of the [SII] line for the periphery of HII region No. 1 (according to the diagram in Fig. 4). For all the other HII regions in Table 2, the line intensities derived from the MPFS
observations were averaged for the whole nebula, so that the largest contribution is from the brightest
region in the corresponding line.

Table 2 demonstrates that, when taken together, the three sets of observations (our MPFS and longslit observations and the data of [2]) cover a wide spectral range and are in good mutual agreement.
The low [NII]/H$\alpha$ line-intensity ratio we measured in the five HII regions is consistent with the
galaxy-mean value of 0.027 given by Martin [6].

\begin{table*}[t!]
\caption{Relative line intensities in the spectra of the  bright HII
regions}
\begin{tabular}{l|c|c|c|c|c|c|c}
\hline
Line & HII 1 & HII 2 & HII 3& HII 4& HII 5 & Giant & Ref. \\
     &       &       &       &       &       &ring & \\
\hline
$[\textrm{OII}]\lambda 3729/[\textrm{OII}]\lambda 3726$       &$1.39\pm 0.11$&$1.43\pm 0.14$&$1.50\pm 0.08$&$1.37\pm 0.10$&$1.80\pm 0.23$&&[2]\\
$[\textrm{OII}](3726+3729$)/H$\beta $&$0.59\pm 0.02$&$2.26\pm 0.11$&$1.18\pm 0.03$&$1.71\pm 0.06$&$2.20\pm 0.12$&&[2]\\
$[\textrm{OIII}]\lambda 5007/$H$\beta $  &$3.89\pm 0.01$&$1.65\pm 0.03$&$2.68\pm 0.01$&$2.59\pm 0.03$&$1.28\pm 0.03$&             &[2]\\
$[\textrm{OIII}]\lambda 5007/$H$\beta $  &$3.43\pm 0.05$&$1.72\pm 0.02$&$2.85\pm 0.03$&$2.78\pm 0.03$&$1.69\pm 0.01$&             & *\\
$[\textrm{OIII}]\lambda 5007/$H$\beta $  &$3.86\pm 0.02$&$1.79\pm 0.04$&$2.85\pm 0.02$&$2.82\pm 0.04$&$1.63\pm 0.07$& $1.32\pm 0.08$& **\\
$[\textrm{OIII}]\lambda 4363/$H$\beta $  &$0.064\pm 0.002$&          &$0.049\pm 0.006$&$0.063\pm 0.009$&        &               &[2]\\
$[\textrm{OI}]\lambda 6300/$H$\beta $&$0.011\pm 0.003$&$0.04\pm 0.01$&$0.044\pm 0.005$&   &          &      &*\\
$[\textrm{SII}]\lambda (6717+6731$)/H$\alpha $&$0.066\pm 0.001$&$0.137\pm 0.001$&$0.123\pm 0.001$&$0.096\pm 0.001$&$0.137\pm 0.003$&    & *\\
$[\textrm{SII}]\lambda (6717+6731$)/H$\alpha $&$0.051\pm 0.001$&$0.140\pm 0.008$&$0.110\pm 0.003$&$0.092\pm 0.005$&$0.137\pm 0.008$&$0.162\pm 0.015$& **\\
$[\textrm{SII}]\lambda 6717/$H$\alpha $ &$0.028\pm 0.001$&\phantom{0}$0.07\pm 0.012$&\phantom{0}$0.06\pm 0.003$&$0.056\pm 0.005$&$0.076\pm 0.09$\phantom{0}&$0.086\pm 0.015$& **\\
$[\textrm{NII}]\lambda (6548+6583$)/H$\alpha $&$0.024\pm 0.001$&$0.032\pm 0.001$&$0.024\pm 0.001$&$0.025\pm 0.003$&$0.038\pm 0.002$& &*\\
$[\textrm{NII}]\lambda 6583/$H$\alpha $   &$0.013\pm 0.001$&$0.027\pm 0.001$&$0.022\pm 0.001$&$0.021\pm 0.001$&$0.034\pm 0.002$& &*\\
H$\gamma $/H$\beta $     &$0.44\pm 0.01$& $0.50\pm 0.01$&$0.42\pm 0.01$&$0.47\pm 0.01$&$0.51\pm 0.02$&$0.00\pm 0.00$&[2]\\
H$\beta $/H$\alpha $     &$0.313\pm 0.002$&$0.304\pm 0.005$&$0.307\pm 0.002$&$0.309\pm 0.004$&$0.255\pm 0.01$\phantom{0}&$0.384\pm 0.013$&**\\
\hline
\multicolumn{8}{l}{$^{~~*}$  --- our data from longslit spectra.}\\
\multicolumn{8}{l}{$^{**}$ --- our MPFS data.}\\
\end{tabular}
\end{table*}

\begin{table*}[t!]
\caption{Extinction and gas parameters in the giant HII regions of  VII~Zw~403.}
\begin{tabular}{l|c|c|c|c|c}
\hline
\multicolumn{1}{c|}{Parameter}     & HII No.~1  & HII No.~2  & HII No.~3 & HII No.~4   & HII No.~5   \\
\hline
$c$        & 0.165   & 0.035   & 0.26   &  0.09    & 0.2   \\
$N_e({\textrm{SII}})$,~cm$^{-3}$ &   57    &   66    &     57 &       110&    --  \\
$T_e({\textrm{SII}})$, K &  10\,000  &   10\,000 &     10\,000&     10\,000&  10\,000 \\
$T_e({\textrm{OIII}})$, K &  14\,500  &    --   &     14\,900&     16\,200&     -- \\
\hline
\end{tabular}  
\end{table*}

\subsection{HII region No. 1: variation of relative line
intensities with distance from the center}

A trend of the relative [SII]/H$\alpha$ line intensity increasing and
[OIII]/H$\beta$ decreasing from the centers of the HII regions to their edges
can be noted in Fig.~3. We considered this effect for the results of our MPFS
observations of the brightest region No.~1.

Figure~4a displays the dependence of the  [SII]$\lambda(6717+6731)$/H$\alpha$ line-intensity ratio and the distance $R$ from the center of HII region No.~1.  We averaged the relative intensities in annular zones at the same distances from the center, the vertical line in the figure corresponds to the boundary
of  shell No.~1 from the HST H$\alpha$ image. A similar dependence on the
distance from the center of the HII region is shown in Fig. 4b for the [OIII]$\lambda5007$/\Hb line-intensity ratio.

The increase in the [SII]/\Ha and decrease in the [OIII]/\Hb line intensities with distance from the center
detected in HII region No. 1 is simple to understand: this trend could indicate that the gas is in a
higher ionization state at the center of the HII region, near the compact OB association, than towards its
edges. In particular, lines of sulfur in higher ionization states than SII dominate in the central region, while virtually all the sulfur at the outer edge of the HII region may be in the form SII. Similarly, the oxygen
at the center is predominantly in the form OIII, and its degree of ionization is lower towards the periphery
of the HII region.

Quantitative estimates of this effect are hindered by the insufficient angular resolution of our observations.
The HST \Ha images show that HII region No. 1 has a shell structure that is narrower than the
angular resolution of our MPFS observations.

Generally speaking, the enhancement of the [SII] lines at the periphery of HII region No. 1 could also
be partially due to a contribution from gas emission behind the shock front. Recall that we detected weak
high-velocity rings in this particular HII region in [1]. However, this contribution is not large: the intensity of the wings detected at velocities up to $300- 500 \km$ is 2-5\% of the maximum in the \Ha and [OIII] lines.

\subsection{Analysis of the HII-region parameters from the measured emission lines}

We used the emission-line intensity ratios from Table~2 to estimate the parameters of the gas in the
giant HII regions and to obtain very rough estimates of the corresponding abundances of oxygen, sulfur, and nitrogen. We estimated the electron densities using the mean intensity ratio  of the [SII]$\lambda6717,6731$\AA\, doublet lines for each  of the bright HII regions.

The electron temperature in the region of [OIII] emission was derived from the intensity ratio of the
auroral and nebular [OIII] lines; the intensities of the $\lambda4363$\AA\, auroral line, quoted in Table~2, were taken from [2]. Unfortunately, the $\lambda4363$\AA\, line was not measured in HII regions Nos. 2 and 5. For the region of emission in the [SII] lines, we adopted the electron
temperature $T_e =10\,000$~K, which is generally characteristic of low-excitation zones.

The parameters of the gaseous components in the HII regions are collected in Table 3. To estimate the
extinction, we used the intensity ratios of the \Hb and \Ha lines from  MPFS data, as well as
measurements of the H$\gamma$ to \Hb line-intensity ratios from Lynds et al. [2]. The two extinction estimates coincide for HII region No. 1; for the other HII regions, Table~3 presents average values for the two studies.

The mean electron density of the gas in the studied HII regions measured from the [SII] lines is $60-70$cm$^{-3}$, in full agreement with the estimates from the [OII] lines presented in [2].

We used the observations of HII region No. 1 to estimate the gas densities at the periphery of
this region, $n_e = 120$~cm$^{-3}$, and at its center, $n_e=30$~cm$^{-3}$. These results agree with the estimate [2] of the mean density for HII region No. 1, $n_e = 50$~cm$^{-3}$, based on measurements of [OII] lines.

\begin{figure}[t!]
\centerline{\includegraphics[width=8 cm]{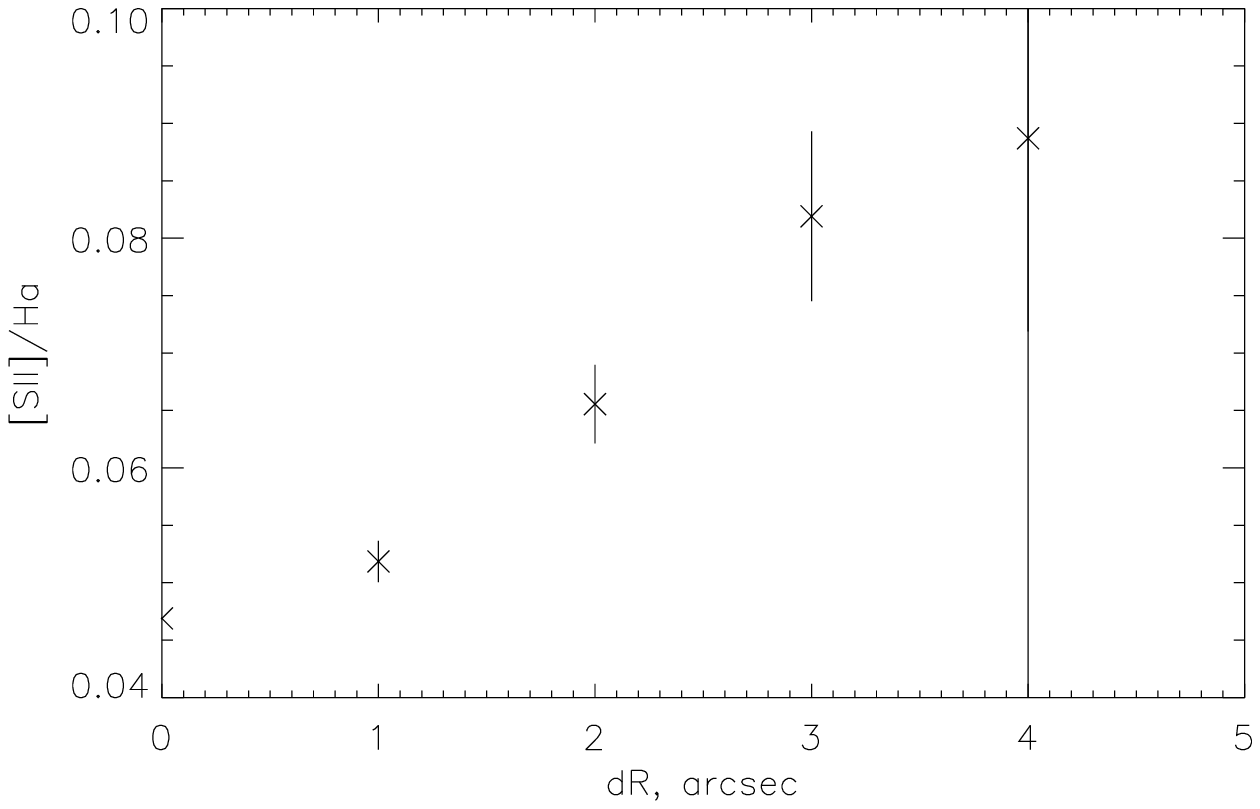}}
\centerline{\includegraphics[width=8 cm]{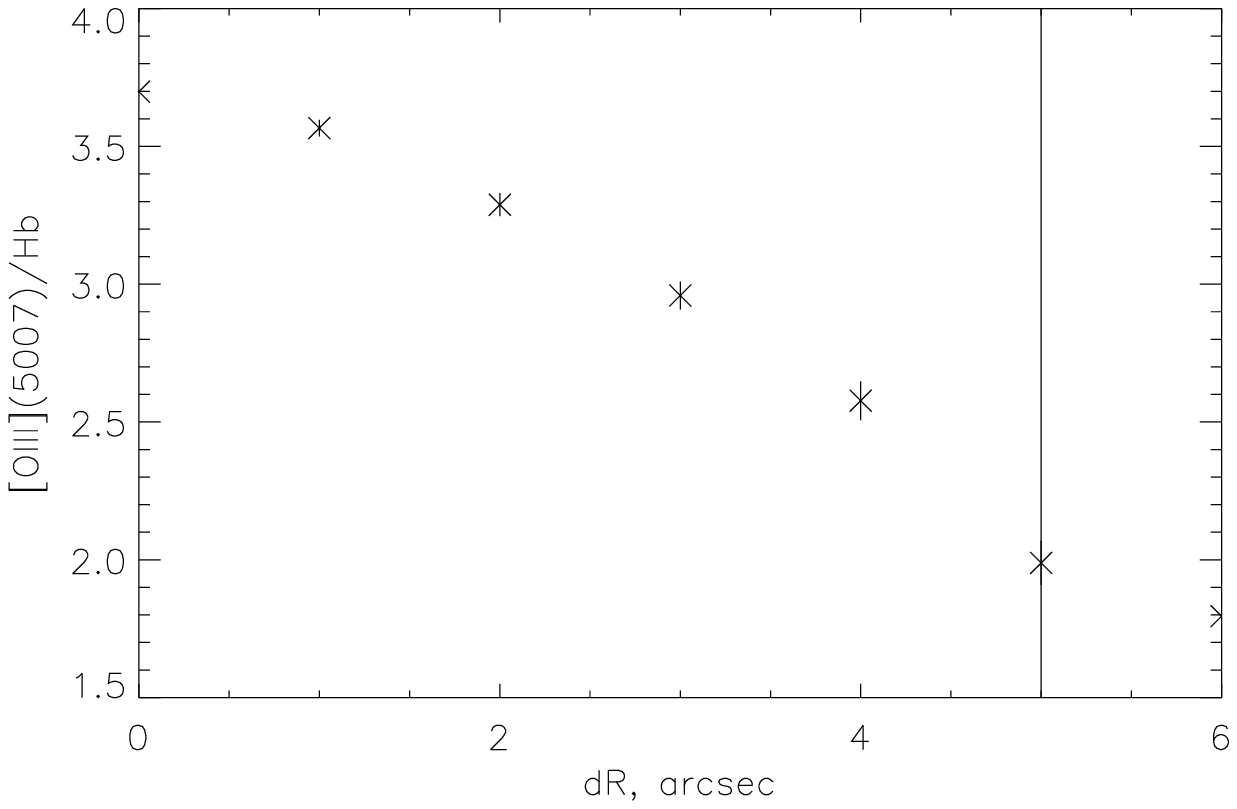}}
\caption{Distribution  of the relative line intensities, for  (a) [SII]$\lambda(6717$+6731)/\Ha and (b) [OIII]$\lambda500$7/H$\beta$ line-intensity ratios versus distance $R$ from the center of region No. 1. The vertical line is the boundary of shell No. 1 from the HST \Ha image.}
\end{figure}

\begin{figure}[t!]
\centerline{\includegraphics[width=8 cm]{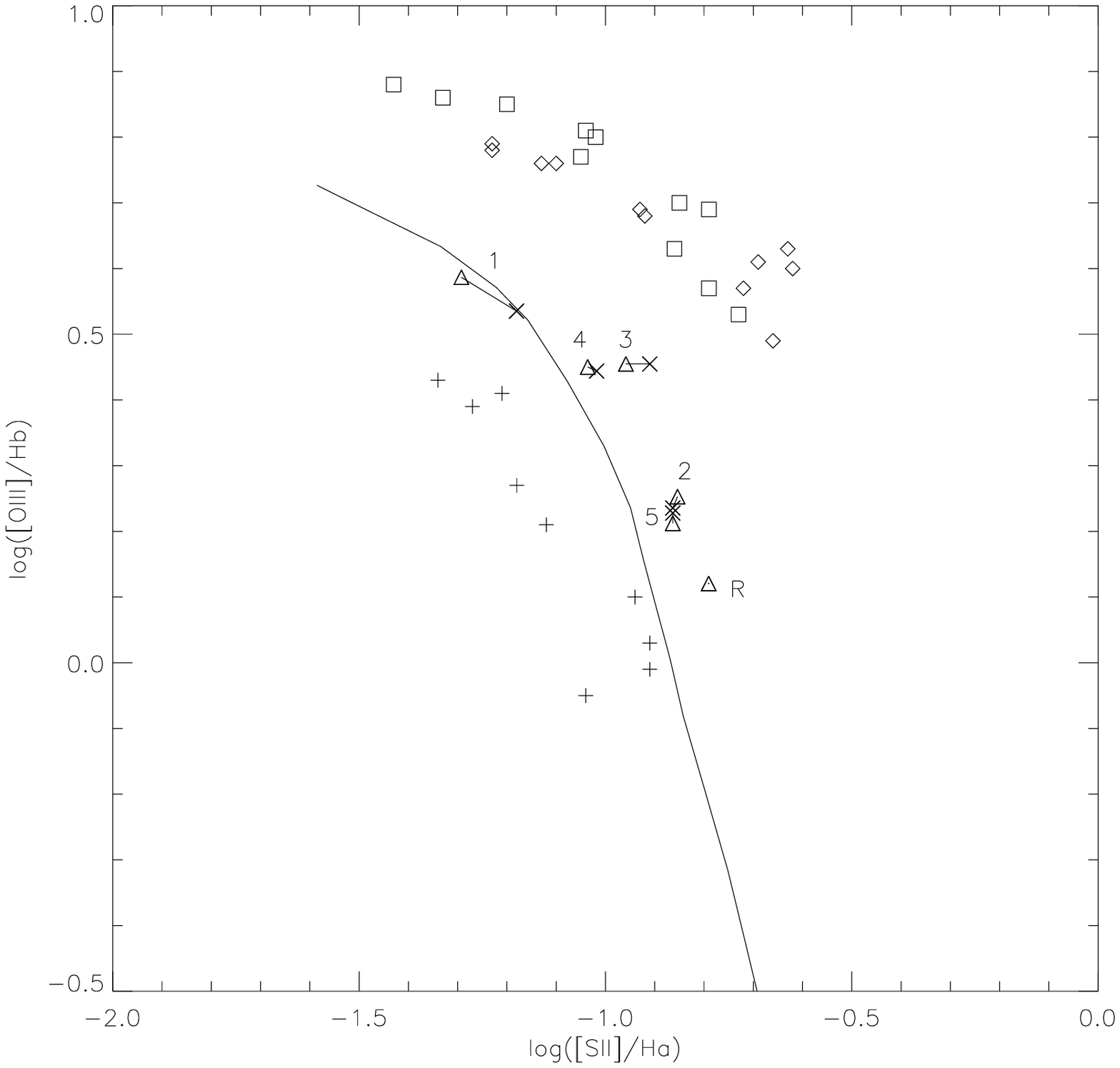}}
\centerline{\includegraphics[width=8 cm]{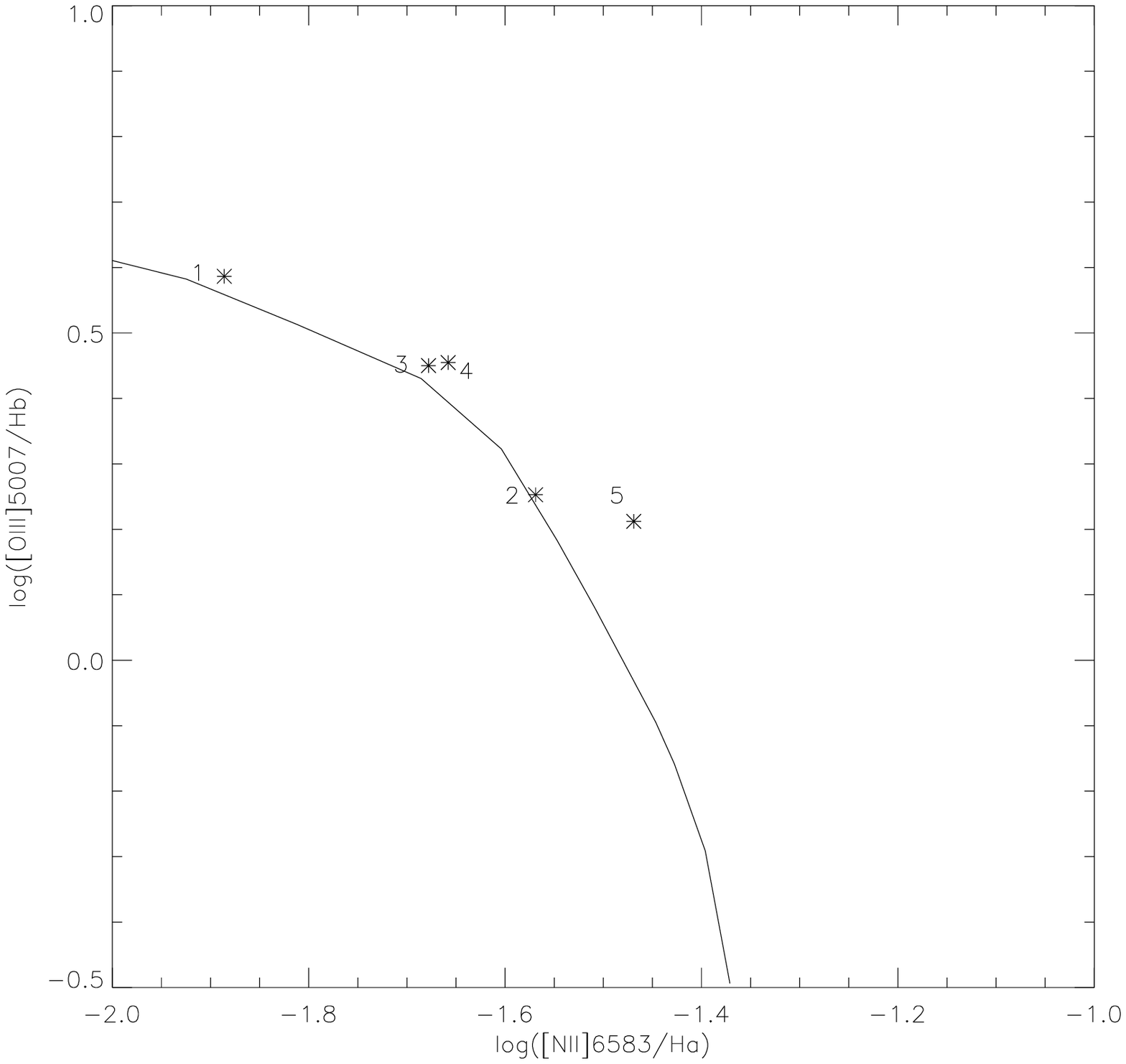}}
\caption{Line intensity ratios from  the MPFS (triangles) and longslit (crosses) data in the (a) [OIII]$\lambda5007$/\Hb--[SII]$\lambda(6717$+6731)/\Ha  and (b)
[OIII]$\lambda5007$/\Hb--[NII]$\lambda6583$/\Ha planes. The five bright HII regions are marked with their numbers,  and the brightest part of the giant ring with the letter $R$. The solid curves
are diagnostic relations derived by Dopita et al.~[16]. Diagram~(a) also shows  measurements for HII regions in the dwarf galaxies I~Zw~18 (pluses), II~Zw~40 (squares), and NGC~4861 (diamonds) from
Martin~[6]. }
\end{figure}

Our estimate, based on the [OIII] lines, of the electron temperature near the center of HII region
No. 1, in the region of the highest intensity of doubly-ionized oxygen, is 14\,500 K, in full agreement with the data of Lynds et al. [2], who obtained $14\,000\pm   200$~K K using the same method.

Our measured line-intensity ratios for the galactic five brightest HII regions and the brightest
part of the giant ring are plotted in Fig. 5 in the [OIII]$\lambda5007$/\Hb--[SII]$\lambda(6717$+6731)/\Ha  and
[OIII]$\lambda5007$/\Hb--[NII]$\lambda6583$/\Ha planes. Figure~5a
also shows the results for HII regions and diffuse ionized gas in other dwarf galaxies from [6]. We
can see that the relative intensities of these lines in VII~Zw~403 are typical of HII regions in dwarf
galaxies.

Figure~5 also presents the diagnostic relations derived by Dopita et al. [16]. We do not use these
diagnostic relations to estimate the gas metallicity in the galaxy for two reasons. First, the computations
of [16] assumed a mass for the central cluster $M\geq 3\times 10^{3} $~$M_{\odot}$, while the compact associations in VII~Zw~403 have lower masses. Second, the models were computed for a single burst of star formation, with the burst results traced for several Myr, while previous star-formation bursts can also be of crucial importance for the abundances of heavy elements in BCD galaxies (see below).

We can see in Fig.~5 that our observations of the five brightest HII regions in VII~Zw~403 are
consistent with the expected `ionization sequence'. In fact, HII regions Nos. 1 and 4 are related to
the richest compact OB associations, Nos. 1 and 5. According to [2], OB association No. 1 contains
25 blue stars with absolute MF555W magnitudes brighter than  $-4^m$ (main-sequence O stars; 27 after
correction for incompleteness) and two type 1b red giants; the luminosity of the surrounding shell No. 1
($L({\textrm{H}}\alpha)= 5\times10^{50}$ photons/s) can be provided by these stars. Association No. 5 contains 28 stars with MF555W brighter than $-4^m$ (blue main-sequence stars and blue supergiants); the luminosity of the corresponding shell No. 4 ($L({\textrm{H}}\alpha)=2.8\times 10^{50} $ photons/s) is about 60\% of the ionizing radiation from these stars. HII region No. 2 is related to the `central' association (see Fig. 11 in [2]), and HII regions Nos. 3 and 5 to OB associations Nos. 4 and 6, respectively. Associations Nos. 4 and 6 are much poorer than No.~1 and No.~5 and contain nine and four blue luminous stars, respectively.

The position of the brightest region of the giant ring in the lowest part of the ionization sequence
appears to correspond to the emission of diffuse ionized gas (as is observed in other galaxies analyzed by
Martin [6]).

We attempted to use the measured line intensities to estimate the relative abundances of oxygen, sulfur,
and nitrogen in the giant HII regions, understanding that the available data can yield only a very rough
approximation.

Assuming that, as was demonstrated above for HII region No. 1, the oxygen near the center of
each of the envelopes is in the state OIII, we can find the total oxygen abundance, assuming that
 OIII/HII $\approx$ O/H. We also assumed that the highest [SII]$\lambda(6717+6731)/$H$\alpha$ and
[NII]$\lambda(6548+6583)/$H$\alpha$ ratios represented the total sulfur and nitrogen abundances,
i.e. SII/HII~$\approx$~S/H and NII/HII~$\approx$~N/H. We derived the ion abundances of doubly-ionized
oxygen and singly-ionized sulfur and nitrogen using the `Five-Level' program of De Robertis et al. [17].

The resulting abundances of O, S , N in the galactic bright HII regions derived in this rough
approximation are listed in Table 4. We can see that, on average for all five HII regions, the logarithmic
oxygen abundance is  $12+\log({\textrm{O}}/{\textrm{H}}) = 7.4{-}7.6$, indicating
an oxygen deficiency by a factor of 16--30 compared to the Sun. Sulfur is also relatively underabundant,
by a factor of 13--30. The mean relative abundance of nitrogen, $12+\log({\textrm{N}}/{\textrm{H}}) = 6.0-6.2$,
corresponds to a deficiency by a factor of 80--120. The N/O relative abundance was found to be 0.03--0.04,
also unusually low.

\begin{figure}[t!]
\includegraphics[scale=1.0]{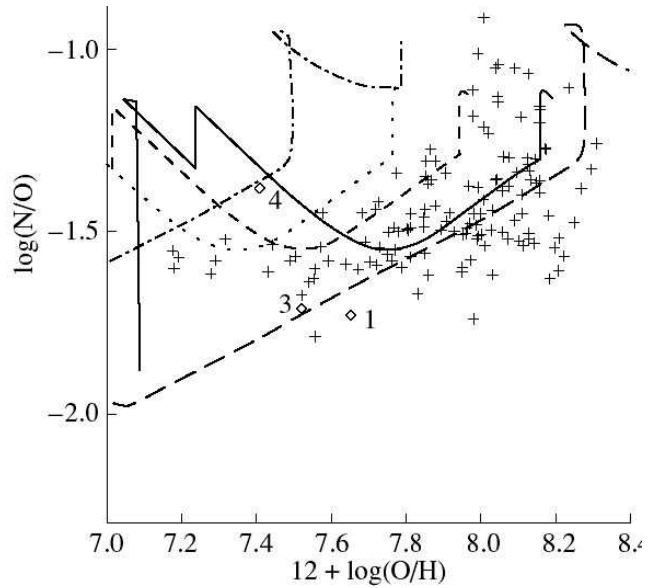}
\caption{Relative nitrogen and oxygen abundances derived for HII regions Nos. 1, 3, and 4 in VII~Zw~403 and for other BCD and HII galaxies from [24, 25] in the $(12+\log ({\textrm{O}}/{\textrm{H}})) {-}  \log
({\textrm{N}}/{\textrm{O}})$ plane. The diagnostic diagrams suggested in [23] for four bursts
of star formation are shown: the dotted curve corresponds to the star-formation rate $\nu = 0.2
\textrm{~Gyr}^{-1}$, the short-dashed
curve to $\nu = 0.3\textrm{~Gyr}^{-1}$, and the solid curve to $\nu = 0.3\textrm{~Gyr}^{-1}$. Also plotted are curves corresponding to two star-formation bursts from [20]: the dot-dashed curve corresponds to $\nu = 1 \textrm{~Gyr}^{-1}$, and the long-dash curve to
$\nu = 10 \textrm{~Gyr}^{-1}$}
\end{figure}

\begin{table*}[t!]
\caption{The relative abundances of oxygen, sulfur, and nitrogen in the giant
HII regions of the galaxy VII~Zw~403 compared to the Sun, in the $12+\log ({\textrm{X}}/{\textrm{H}})$ scale.}
\begin{tabular}{l|c|c|c|c|c|c}
\hline
\multicolumn{1}{c|}{Element}  & HII No.~1  & HII No.~2  & HII No.~3 & HII No.~4  & HII No.~5  & Sun  \\
\hline
   O      &  7.65   &         &   7.52 &    7.41  &         &  8.93 \\
   S     &   5.55   &   5.93  &  5.86 &  5.80  &    5.93 &  7.18 \\
   N     &  5.91   &   6.08  &   5.80 &   6.04  &    6.22 &  8.05 \\
$\log ({\textrm{N}}/{\textrm{O}})$ &  --1.73  &         &   --1.71&  --1.38  &         &  \\
$\log ({\textrm{S}}/{\textrm{O}})$ &  --1.82  &         &   --1.66&  --1.62  &         & \\
 \hline
\end{tabular}
\end{table*}

\section{Discussion}

Obviously, our spectral range is too narrow to provide accurate determinations of the galactic metallicity,
and the above assumptions are very rough. However, the oxygen abundances we derived for the
HII regions in the galaxy VII~Zw~403 (with both the temperature and density being determined from
lines for this element) are in very good agreement with the data of other studies. Martin [6] found the
average value $12+\log ({\textrm{O}}/{\textrm{H}}) = 7.58   \pm  0.01$ for the
galaxy. Introducing corrections for unobserved ionization states, Izotov et al. [18] found the galaxy average
to be $12+\log ({\textrm{O}}/{\textrm{H}}) = 7.69  \pm  0.001$ Schulte-Ladbeck et al. [19] found the metallicity of the galactic gas to be $Z = (0.05 {-}0.06) $~$Z_{\odot}$.

Our assumption that oxygen near the center of an HII region is predominantly in the state OIII is
confirmed by a comparison with the data in Table 4 of Izotov et al. [18], where the sum OII+OIII is given
for the observed region of VII~Zw~403, resulting in the same oxygen abundance as in our study.

The very low abundance of nitrogen relative to oxygen in the giant HII regions of VII~Zw~403 can
be explained in the models for the chemical evolution of BCD galaxies suggested by Brandamante et
al. [20], Recchi et al. [21, 22], and Lafranchi and Matteucci [23] (see also references in [23]). According to
these last authors, the observed heavy-element abundances in BCD galaxies can be understood in terms of
the chemical evolution of these galaxies if there were several (two to seven) relatively short bursts of star
formation separated by long `quiescent' periods. Figure 6 displays the diagnostic diagrams in
the $(12+\log ({\textrm{O}}/{\textrm{H}})) - \log ({\textrm{N}}/{\textrm{O}})$ plane suggested
in [23] for a model with four star-formation bursts at epochs corresponding to galaxy ages of 1, 11, 13, and 13.98 Gyr, with the burst durations being 20, 10, 200, and 20 Myr, respectively. The different curves correspond to different star-formation rates. Also shown are curves corresponding to two short bursts separated by a quiescent period for two extreme star-formation rates from Brandamante et al. [20].

Figure~6 shows the relative abundances we derived for HII regions Nos. 1, 3, and 4 using the parameters
presented in Table 3 together with the results for other BCD and HII galaxies according to the data
from Izotov and Thuan [24] and from Kobulnicky and Skillman [25].We see that the positions of HII regions
Nos. 1, 3, and 4 in the diagnostic diagram agree with regions occupied by other galaxies of the same
type and are consistent with the theoretical model of Lafranchi and Matteucci [23] with four star-formation
bursts. Those authors note that the lowest N/O ratio is achieved during the third burst, explained by an
increase in the oxygen abundance due to the injection of oxygen into the interstellar gas during massive
type II supernovae explosions. (The injection of nitrogen `lags behind' that of oxygen, since it is brought
about by winds from lower-mass stars.)

It follows from Fig. 6 that, in the model with four star-formation bursts, our observations are consistent
with the theoretical curves corresponding to the star-forming efficiency $\nu\approx0.2{-}0.3\textrm{~Gyr}^{-1}$. This in no way contradicts any observations of VII~Zw~403. Using the formula of Kennicutt [26]
for estimating the current star-formation rate from a galactic \Ha luminosity, we find from the value
$L({\textrm{H}}\alpha) = (1.49{-}1.86)\times 10^{39}$~erg~s$^{-1}$ we derived in [1] the star-formation rate $(0.012{-}0.015)$~$M_{\odot}$/yr for masses between 0.15 and 100~$M_{\odot}$. (Our measurements
did not confirm the results of Silich et al. [7], who gave a much higher \Ha luminosity for the galaxy
(by a factor of 20), though our results are in good agreement with earlier measurements [2, 27].)

Given this star-formation rate and a total mass of gas in the galaxy near $10^{8}\,M_{\odot}$ (the HI mass estimated by Thuan and Martin [28] corrected for the new distance, 4.5 Mpc [2]), a gas-depletion time of
about 8 Gyr corresponds to $\nu =0.125$Gyr${-1}$. This is probably an upper limit; according to Bicker
and Fritze-van Alvensleben [29], the use of the relation from [26], derived for the solar chemical composition,
overestimates the star-formation rates of low metallicity galaxies.

The previous burst of star formation, which was responsible for the observed relative N/O abundance
in the model of Lafranchi and Matteucci [23], was appreciably stronger than the current one. According
to the model of [23], which is in good agreement with our observations, the last star-formation burst in
VII~Zw~403 commenced some 20 Myr ago. A detailed comparison of the galactic chemical
evolution with the abundances of heavy elements observed in VII~Zw~403 is beyond the scope of this
study. We note only that multicolor stellar photometry with the HST suggests that several star-formation
episodes with various intensities have occurred in the galaxy [2, 3].

Three epochs of active star formation in VII~Zw~403 were identified in [3]. The first, identified from stars on the red-giant branch, occurred 3--10 Gyr ago, or possibly even earlier; the second burst, identified from stars on the asymptotic giant branch (AGB), occurred about 4 Gyr ago. The age of the third burst, which was responsible for the young stars and HII regions we observe today, does not exceed one Gyr.

Lynds et al. [2] also noted the rich population of AGB stars with ages of 1-5 Gyr, which is unusual for dwarf galaxies; the strongest starformation burst in VII~Zw~403 happened between 800 and 600 Myr ago, possibly even 1.2 Gyr ago. The stars of this population dominate in
the galaxy. The ages of six compact OB associations in the galactic central parts are about 4-6 Myr. (This estimate is fairly rough due to the small number of stars in the color-magnitude diagrams.)
In addition to these compact associations, blue luminous stars are present in the galactic entire central
field, out to a distance of one kpc. In particular, two stellar populations with ages of 5 and 10 Myr, or possibly somewhat more, are observed in the less compact stellar groups of mixed age, designated
`center', `north', and `south' in Fig. 11 of [2]. Thus, the stellar population of the galaxy exhibits signs of
recent star formation that occurred in two bursts, about 5 Myr ago and about 10 Myr
ago or somewhat earlier, with the strongest burst occurring 600-800 Myr ago (possibly about
1.2 Gyr ago), in complete agreement with the model of Lanfranchi and Matteucci [23] that provides
the best approximation to our observational results for VII~Zw~403.

Our study is the first to estimate the relative oxygen, sulfur, and nitrogen abundances for individual
HII regions in VII~Zw~403; results published earlier give the average chemical composition for
the galaxy. The differences of the relative abundances $12+\log ({\textrm{O}}/{\textrm{H}})$) in HII regions Nos. 1, 3, and 4 of VII~Zw~403 we have found (from 7.65 to 7.41; Table 4) are similar to the results obtained
by Papaderos et al. [30] for the galaxy SBS~0335-052E. Both galaxies have similar \Ha morphologies;
in SBS 0335-052E, an O/H gradient between 7.20 and 7.31 on a scale less than one kpc was discovered.
As for VII~Zw~403, the highest oxygen abundance is observed in the brightest HII region of the galaxy,
around the youngest cluster.

\begin{acknowledgements}
This study was supported by the Russian Foundation for Basic Research (project codes 04-02-16042 and 05-02-16454). The study is based on observations obtained at the 6-m telescope of the Special
Astrophysical Observatory (Russian Academy of Science), which is supported by the Ministry of Education and Science of the Russian Federation (registration number 01-43). The authors are grateful
to O.K. Sil'chenko for helpful discussions.
\end{acknowledgements}

\textit{Translated by N. Samus'}

\end{document}